# Inhomogeneous maps: the basic theorems and some applications

Georgy P. Karev

National Center for Biotechnology Information, National Institutes of Health, Bethesda, MD 20894, USA. Email: karev@ncbi.nlm.nih.gov

**Abstract.** Non-linear maps can possess various dynamical behaviors varying from stable steady states and cycles to chaotic oscillations. Most models assume that individuals within a given population are identical ignoring the fundamental role of variation. Here we develop a theory of inhomogeneous maps and apply the general approach to modeling heterogeneous populations with discrete evolutionary time step. We show that the behavior of the inhomogeneous maps may possess complex transition regimes, which depends both on the mean and the variance of the initial parameter distribution. The examples of inhomogeneous models are discussed.

### 1. Statement of the problem and basic Theorem

Let us assume that a population consists of individuals, each of those is characterized by its own parameter value $\mathbf{a} = (a_1,...,a_k)$. These parameter values can take any particular value from set A. Let $n_t(\mathbf{a})$ be the density of the population at the moment $t$. Then the number of individuals having parameter values in set $\widetilde{A} \subseteq A$ is given by $\widetilde{N}_t = \int_{\widetilde{A}} n_t(\mathbf{a})d\mathbf{a}$, and the total population size is $N_t = \int_A n_t(\mathbf{a})d\mathbf{a}$.

The theory of inhomogeneous models of populations with continuous time was developed earlier (see, e.g., [2], [3]). Here we study the inhomogeneous model of population dynamics with discrete time of the form

$$n_{t+1}(\mathbf{a}) = W(N_t, \mathbf{a})n_t(\mathbf{a}), \quad N_t = \int_A n_t(\mathbf{a})d\mathbf{a} \tag{1}$$

where $W \geq 0$ is the reproduction rate (fitness); we assume that the reproduction rate depends on the specific parameter value **a** and the total size of the population, $N_t$, but does not depend on the particular densities $n_t$.

Let us denote $p_t(\mathbf{a}) = n_t(\mathbf{a})/N_t$ the current probability density of the vector-parameter **a** at the moment $t$. We have a probability space $(A, \mathbf{P}_t)$ where the probability $\mathbf{P}_t$ has the density $p_t(\mathbf{a})$, and model (1) defines a transformation of the initial probability density $p_0(\mathbf{a})$ with time. Below we show that problem (1) can be reduced to a non-autonomous map on $I \subseteq \mathbf{R}^1$ under supposition that the reproduction rate has the form $W(N, \mathbf{a}) = f(\mathbf{a})g(N)$, so that the model takes the form

$$n_{t+1}(\mathbf{a}) = n_t(\mathbf{a})f(\mathbf{a})g(N_t), \quad N_t = \int_A n_t(\mathbf{a})d\mathbf{a}, \tag{2}$$

for the given initial density $n_0(\mathbf{a})$.

**Theorem 1.** *Let $p_0(\mathbf{a})$ be the density of the initial probability distribution of the vector-parameter **a** for inhomogeneous map (2). Then*

i) *The population size $N_t$ satisfies the recurrence relation $N_{t+1} = E_t[f]N_t g(N_t)$;* (3)

ii) *The current mean of $f$ can be computed by the formula $E_t[f] = E_0[f^{t+1}]/E_0[f^t]$;* (4)

iii) *The density of the current distribution is $p_t(\mathbf{a}) = p_0(\mathbf{a})f^t(\mathbf{a})/E_0[f^t]$.* (5)

Proof.

Rewriting the first equation in (2) as $n_{t+1}(\mathbf{a})/n_t(\mathbf{a}) = f(\mathbf{a})g(N_t)$, we obtain $n_t(\mathbf{a}) = n_0(\mathbf{a})f^t(\mathbf{a})G_{t-1}$, where $G_t = g(N_0) \cdot \ldots \cdot g(N_t)$. Then

$$N_t = \int_A n_t(\mathbf{a})d\mathbf{a} = \int_A n_0(\mathbf{a})f^t(\mathbf{a})G_{t-1}d\mathbf{a} = N_0 E_0[f^t]G_{t-1},$$

where $E_0[f^k] = \int_A f^k(\mathbf{a})p_0(\mathbf{a})d\mathbf{a}$. So one has $p_t(\mathbf{a}) = \dfrac{n_t(\mathbf{a})}{N_t} = p_0(\mathbf{a})\dfrac{f^t(\mathbf{a})}{E_0[f^t]}$.

Integrating over **a** the equation $n_{t+1}(\mathbf{a}) = f(\mathbf{a})p_t(\mathbf{a})N_t g(N_t)$ implies $N_{t+1} = E_t[f]N_t g(N_t)$ where $E_t[f] = \int_A f(\mathbf{a})p_t(\mathbf{a})d\mathbf{a}$. Next, $E_t[f] = \dfrac{1}{E_0[f^t]}\int_A f^{t+1}(\mathbf{a})p_0(\mathbf{a})d\mathbf{a} = \dfrac{E_0[f^{t+1}]}{E_0[f^t]}$. Q.E.D.

## 2. Dynamics of the parameter distributions

The problem of the evolution of parameter distribution due to inhomogeneous model (1) is of especial interest. Formally assertion iii) of Theorem 1 contains complete description of $p_t(\mathbf{a})$. Roughly, the density $p_t(\mathbf{a})$ tends to 0 if $f(\mathbf{a})<1$ and tends to $\infty$ if $f(\mathbf{a})>1$ at $t \to \infty$. The following proposition (immediately following from Theorem 1) gives some additional useful relations.

**Proposition 1.** $p_{t+1}(\mathbf{a}) = p_t(\mathbf{a})f(\mathbf{a})/ E_t[f]$; $p_t(\mathbf{a}_1)/ p_t(\mathbf{a}_2) = p_0(\mathbf{a}_1)/ p_0(\mathbf{a}_2) [f(\mathbf{a}_1)/f(\mathbf{a}_2)]^t$.

**Corollary 1.** If $f(\mathbf{a}_1) < f(\mathbf{a}_2)$ and $p_0(\mathbf{a}_2) > 0$, then $p_t(\mathbf{a}_1)/ p_t(\mathbf{a}_2) \to 0$ at $t \to \infty$.

The density independent component of the reproduction rate $f(\mathbf{a})$ can be considered as a random variable on the probability space $(A, \mathbf{P}_t)$. Let $p_f(t;x)$ be the probability density function (pdf) of this r.v. $f(\mathbf{a})$; then $E_t[f] = \int_{-\infty}^{\infty} x p_f(t;x) dx$ and $\mathrm{Var}_t[f] = E_t[f^2] - (E_t[f])^2 =$

$E_0[f^{t+2}]/ E_0[f^t] - (E_0[f^{t+1}]/E_0[f^t])^2$ according to Theorem 1, ii).

**Proposition 2.** $E_{t+1}[f] = E_t[f^2]/ E_t[f]$, $E_{t+1}[f]/E_t[f] = 1 + \mathrm{Var}_t[f]/ E^2_t[f]$.

**Corollary 2.** $\Delta E_t[f] = \mathrm{Var}_t[f]/ E_t[f]$.

Hence, $E_{t+1}[f] > E_t[f]$ for all $t$ and $E_{t+1}[f] = E_t[f]$ if and only if $\mathrm{Var}_t[f]=0$, i.e. if $f(\mathbf{a})$=const for almost all $\mathbf{a}$ over the probability $\mathbf{P}_t$. Remark, that this corollary is a version of the Fisher' fundamental theorem of natural selection within a framework of considered model (2).

Next, let us explore the evolution of the distribution of $f(\mathbf{a})$ in detail.

**Theorem 2.** *Let the initial pdf of $f(\mathbf{a})$ be*

1) *Gamma-distribution with the parameters $(s, k)$, i.e.* $p_f(0;x) = s^k x^{k-1} \exp[-xs]/\Gamma(k)$ *for $x \geq 0$, where $s, k$ are positive, $\Gamma(k)$ is the $\Gamma$-function. Then*

$p_f(t;x) = s^{t+k} x^{t+k-1} \exp[-xs] /\Gamma(k+t)$

*is again the Gamma-distribution with the parameters $s, k+t$; its mean is $E_t[f] = (k+t)/s$ and variance $\mathrm{Var}_t[f] = (k+t)^2/s^2$.*

2) *exponential, i.e.* $p_f(0;x) = s\exp(-sx)$, $s \geq 0$ *is a parameter. Then*

$p_f(t;x) = s\exp(-sx) (sx)^t/t!$

*is the density of Gamma-distribution with the parameters $(s, 1+t)$.*

3) *Beta-distribution with parameters $(\alpha, \beta)$ where $\alpha, \beta$ are positive, i.e.*

$p_f(0;x) = \dfrac{\Gamma(\alpha + \beta)}{\Gamma(\alpha)\Gamma(\beta)} x^{\alpha-1}(1-x)^{\beta-1}$, $0<x<1$. *Then*

$$p_f(t;x) = \frac{\Gamma(\alpha + t + \beta)}{\Gamma(\alpha + t)\Gamma(\beta)} x^{\alpha+t-1} (1-x)^{\beta-1}$$

is again the density of Beta-distribution with parameters $(\alpha+t, \beta)$; its mean is
$E_t[f]=(\alpha+t)/(\alpha+t+\beta)$ and variance $\mathrm{Var}_t[f]= (\alpha+t) \beta/[(\alpha+t+\beta)^2 (\alpha+t+\beta+1)]$.

4) *hyper-exponential, i.e.* $p_f(0;x) = \sum_{k=1}^{m} \alpha_k s_k \exp(-s_k x)$, $x \geq 0$. Then

$$p_f(t;x) = x^t \sum_{k=1}^{m} \alpha_k s_k \exp(-s_k x) / (\sum_{k=1}^{m} \alpha_k t!/s_k^t).$$

5) *log-normal, i.e.* $p_f(0;x) = \dfrac{1}{x\sigma\sqrt{2\pi}} \exp\{-\dfrac{(\ln x - m)^2}{2\sigma^2}\}$, $x>0$. Then

$$p_f(t;x) = \frac{x^{t-1}}{\sigma\sqrt{2\pi}} \exp\{-\frac{(\ln x - m)^2}{2\sigma^2} - \frac{t^2\sigma^2}{2} - tm\}.$$

6) *Pareto distribution with the parameters* $\alpha, x_0$, *i.e.* $p_f(0;x) = \alpha/x_0 (x/x_0)^{-\alpha-1}$, $x>x_0>0$. Then

$$p_f(t;x) = (\alpha-t)/x_0 (x/x_0)^{-\alpha-1+t}$$

is again the Pareto distribution for $t< \alpha$ with the parameters $\alpha-t, x_0$.

7) *Veibull distribution with parameters* $(k,s)$, *i.e.* $p_f(0,x) = ksx^{k-1}\exp(-sx^k)$, $x>0$. Then

$$p_f(t;x) = ks^{1+t/k} x^{t+k-1} \exp(-sx^k)/\Gamma(t/k+1).$$

8) *uniform distribution in the interval* $[0,B]$. Then

$$p_f(t;x) = (t+1) x^t / B^{t+1}.$$

Proof.

1) If $p_f(0;x) = s^k x^{k-1}\exp[-xs]/\Gamma(k)$, then $E_0[f^t]= \Gamma(k+t)/(s^t\Gamma(k))$. According to formula (5),

$$p_f(t;x) = p_f(0,x) x^t/ E_0[f^t] = s^{t+k}x^{t+k-1}\exp[-xs] /\Gamma(k+t),$$

and $p_f(t;x)$ is again the Gamma-distribution with the parameters $s, k+t$.

2) If $p_f(0;x) = s\exp(-sx)$, then $E_0[f^t]=t!/s^t$. Hence,

$$p_f(t;x) = p_f(0;x) x^t/ E_0 [f^t] = s\exp(-sx) (sx)^t/t!.$$

Remark, that under fixed value of $x$ the last formula defines (up to normalized constant $1/s$) the Poissonian distribution over time instants with the parameter $sx$.

3) If $p_f(0;x) = \dfrac{\Gamma(\alpha + \beta)}{\Gamma(\alpha)\Gamma(\beta)} x^{\alpha-1}(1-x)^{\beta-1}$ then $E_0[f^t]= \dfrac{\Gamma(\alpha + t)\Gamma(\alpha + \beta)}{\Gamma(\alpha)\Gamma(\alpha + t + \beta)}$. Hence,

$$p_f(t;x) = p_f(0;x) x^t/ E_0[f^t] = \frac{\Gamma(\alpha + t + \beta)}{\Gamma(\alpha + t)\Gamma(\beta)} x^{\alpha+t-1} (1-x)^{\beta-1}$$

is the *Beta*-distribution with parameters $(\alpha+t, \beta)$.

4) If $p_f(0;x) = \sum_{k=1}^{m} \alpha_k s_k \exp(-s_k x)$, then $E_0[f^t] = \sum_{k=1}^{m} \alpha_k t!/s_k^t$. Hence,

$$p_f(t;x) = p_f(0,x)\, x^t/\, E_0[f^t] = x^t \sum_{k=1}^{m} \alpha_k s_k \exp(-s_k x)/(\sum_{k=1}^{m} \alpha_k t!/s_k^t).$$

5) If $p_f(0;x) = \dfrac{1}{x\sigma\sqrt{2\pi}} \exp\{-\dfrac{(\ln x - m)^2}{2\sigma^2}\}$, then $E_0[f^t] = \exp\{1/2\, t^2 \sigma^2 + tm\}$. Hence,

$$p_f(t;x) = p_f(0,x)\, x^t/\, E_0[f^t] = \dfrac{x^{t-1}}{\sigma\sqrt{2\pi}} \exp\{-\dfrac{(\ln x - m)^2}{2\sigma^2} - \dfrac{t^2\sigma^2}{2} - tm\}.$$

6) If $p_f(0;x) = \alpha/x_0 \,(x/x_0)^{-\alpha-1}$, then $E_0[f^t] = \alpha/(\alpha-t) x_0^t$ for $t < \alpha$. Hence,

$$p_f(t;x) = p_f(0;x)\, x^t/\, E_0[f^t] = (\alpha-t)/x_0\, (x/x_0)^{-\alpha-1+t}.$$

7) If $p_f(0,x) = ks x^{k-1} \exp(-sx^k)$, then $E_0[f^t] = s^{-t/k} \Gamma(t/k+1)$, hence

$$p_f(t;x) = p_f(0;x)\, x^t/\, E_0[f^t] = ks^{1+t/k} x^{t+k-1} \exp(-sx^k)/\Gamma(t/k+1).$$

8) If $p_f(0;x) = 1/B$, then $E_0[f^t] = B^t/(t+1)$, hence

$$p_f(t;x) = p_f(0;x)\, x^t/\, E_0[f^t] = (t+1)\, a^t / B^{t+1}.\ \text{Q.E.D.}$$

Evolution of other initial distributions of the fitness can be explored similarly.

## 3. Examples

**Example A**. *Malthusian model of the population growth*

Inhomogeneous version of the Malthusian model read $n(t+1,\mathbf{a}) = f(\mathbf{a}) n(t,\mathbf{a})$. Even in this simplest case the dynamics of the mean fitness and the evolution of the fitness' distribution over the individuals dramatically depend on the initial distribution of the fitness. Let us consider some important examples.

A1) Let the initial pdf $p_f(0;x)$ of $f(\mathbf{a})$ be the Gamma-distribution with the parameters $s, k$. Then, according Theorem 2, 1) $p_f(t;x)$ is again the Gamma-distribution with the parameters $s$, $k+t$; its mean is $E_t[f] = (k+t)/s$, $\mathrm{Var}_t[f] = (k+t)^2/s^2$, and $N_{t+1} = E_t[f] N_t = (k+t)/s\, N_t$.

Hence in this case the mean fitness increases linearly with time, while $N_t = N_0 \Gamma(k+t)/(s^t \Gamma(k)) \sim N_0 t^k/s^t$; in particular, with $s=1$ the population size increases asymptotically as a power function.

A2) Let $p_f(0;x)$ be the Beta-distribution in interval $[0,B]$. Then $E_t[f] = B(\alpha+t)/(\alpha+t+\beta) \sim B$.

Next, $N_{t+1} = N_t\, E_t[f] = B(\alpha+t)/(\alpha+t+\beta) N_t$, so

$$N_t = N_0 B^t \frac{\Gamma(\alpha+t-1)\Gamma(\alpha+\beta)}{\Gamma(\alpha+\beta+t-1)\Gamma(\alpha)} \sim N_0 B^t \frac{\Gamma(\alpha+\beta)}{\Gamma(\alpha)} t^\beta.$$

Hence, the fate of a population dramatically depends on the value of $B$: if $B \leq 1$, the population goes to extinct, if $B > 1$, the size of the population increases indefinitely. In the case $B = 1$ the mean fitness tends to 1 and one could expect, that the total population size tends to a stable non-zero value in course of time, but actually the population goes to extinct with a power rate.

A3) Let $p_f(0;x)$ be the uniform distribution in the interval $[0,B]$. Then $E_0[a^t] = B^t/(t+1)$, hence

$E_t[f] = B(t+1)/(t+2) \sim B$, and $N_t = N_0 B^t /(t+1)$.

Similar to the previous example, the fate of a population depends on the value of $B$: if $B \leq 1$, the population goes to extinct, if $B > 1$, the size of the population increases indefinitely.

A4) Let $p_f(0;x)$ be the log-normal distribution. Then $E_t[f] = \exp\{1/2\sigma^2 (2t+3) + m\} \sim \exp\{\sigma^2 t\}$ and $\Delta E_t[f] = \exp\{t\sigma^2 + 3/2\sigma^2 + m\}(1 - \exp\{-\sigma^2\})$. The mean fitness increases exponentially with time. Next, $N_{t+1} = N_t E_t[f] = \exp\{1/2\sigma^2 (2t+3) + m\} N_t$, so

$N_t = N_0 \exp\{1/2\sigma^2 (t^2 + 2t) + mt\} \sim N_0 \exp\{1/2\sigma^2 t^2\}$.

**Example B**. The *Ricker' model* is the map of the form

$N(t+1) = N(t) \lambda \exp(-\beta N_t)$, $\lambda, \beta > 0$ are parameters.

Consider the inhomogeneous version of this model with distributed parameter $\lambda$,

$n(t+1,\mathbf{a}) = n(t,\mathbf{a}) F(\mathbf{a}, N_t) = n(t,\mathbf{a}) \lambda_0 f(\mathbf{a}) \exp(-\beta N_t)$

where $\lambda_0$ is the scaling multiplier. Let the initial pdf of $f(\mathbf{a})$, $p_f(0;x)$, be the Gamma-distribution with the parameters $(s, k)$. Then, according to Theorem 1, $p_f(t;x)$ is the Gamma-distribution with parameters $(s, k+t)$, and $E_t[F] = \lambda_0(k+t)/s \exp(-\beta N_t)$. So,

$N_{t+1} = N_t \lambda_0(k+t)/s \exp(-\beta N_t)$.

The coefficient $\lambda_0(k+t)/s$, which determines the dynamics of the Ricker' model increases indefinitely with time and hence after some time moment the population size begins to oscillate with increasing amplitude (according to the theory of the plain Ricker model, see [ ]).

If the parameter $\lambda_0$ is small and/or $s$ is large then the sequence $\{\lambda_0(k+t)/s, t=0,1,...\}$ takes the values close to all bifurcation values of the coefficient $\lambda$ of the plain Ricker' model. It follows from here a notable phenomenon: "almost complete" (with the step $\lambda/s$) sequence of all possible bifurcations of the Ricker' model is realized in frameworks of unique inhomogeneous Ricker model, see Figure 1. The trajectory $\{N_t\}_0^\infty$ in some sense mimics the bifurcation diagram of the plain Ricker' model. The model' evolution goes through different stages with the speed depending on $\lambda_0/s$.

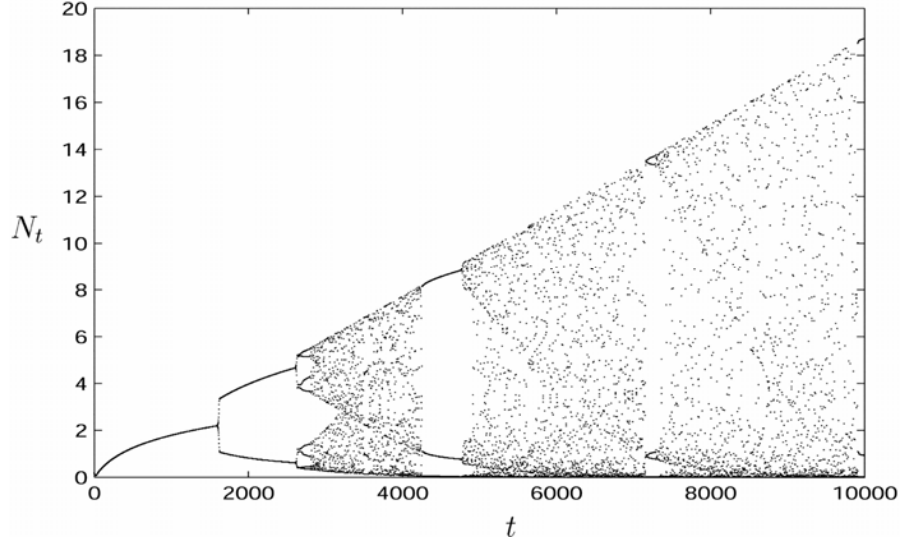

Fig. 1. The trajectory of the inhomogeneous Ricker' model with Gamma-distributed parameter.

We can observe the similar phenomenon for any initial distribution of the parameter with unbounded range of values of $f(\mathbf{a})$. For example, if $p_f(0;x)$, be the log-normal distribution, then corresponding version of inhomogeneous model read $N_{t+1}=N_t \lambda \exp\{1/2\sigma^2 (2t+3)+m-\beta N_t\}$.

The model shows other behavior if the range of values of $f(\mathbf{a})$ is bounded. It is clear (see Corollary 1) that the final dynamics at $t\to\infty$ of models with any initial distribution and bounded range of values of $f(\mathbf{a})$ is determined by the maximal possible value of $f(\mathbf{a})$.

Let, for example, $p_f(0;x)$ be the uniform distribution in the interval [0,1]. Then

$N_{t+1} = N_t \lambda_0(t+1)/(t+2) \exp(-\beta N_t)$.

If $p_f(0;x)$ is the Beta-distribution in [0,1] with parameters $(\alpha,\beta)$, then, as it was shown above, $p_f(t;x)$ is again the Beta-distribution with parameters $(\alpha+t,\beta)$ and hence,

$N_{t+1} = N_t \lambda_0(t+\alpha)/(t+\alpha+\beta) \exp(-\beta N_t)$.

Choosing appropriate value of $\lambda_0$ we will observe as the final dynamics behavior any possible behavior of the model. The following figure 2 illustrates this assertion.

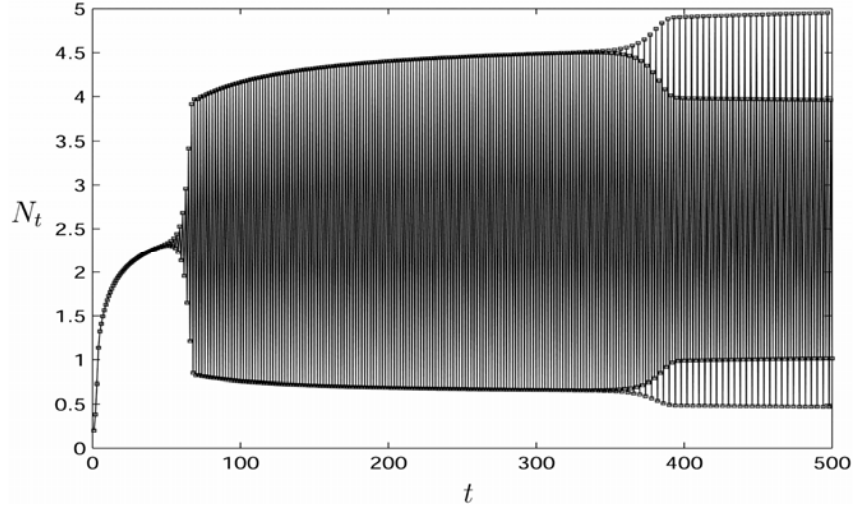

Fig. 2. The trajectory of the inhomogeneous Ricker' model with Beta-distributed parameter ($\lambda_0$=14, $\alpha$=3, $\beta$=20). The final dynamics of the model is 4-cycle.

Inhomogeneous versions of other well-known maps such as logistic, Skellam' model, etc. can be explored the same way.

**Example C. Non-homogeneous model of natural rotifer population**

The mathematical model of zooplankton populations, extracted as deterministic dynamics components from noisy ecological time series and studied systematically in [1], is of the form

$$N_{t+1} = N_t \exp\{-a + 1/N_t - \gamma/N_t^2\}. \tag{6}$$

Here $a$ is the parameter characterizing the environment quality, and $\gamma$ is the species-specific parameter. Assuming that parameter $a$ is distributed, we have the model in the form (3) with $f(a) = \exp\{-a\}$, and $g(N) = \exp\{1/N - \gamma/N^2\}$. Let us assume that the initial distribution of $a$ is a Gamma-distribution with parameters $b, k, s$, $p_0(a) = \dfrac{s^k}{\Gamma(k)}(a-b)^{k-1}\exp\{-(a-b)s\}$. Then $E_0[f^t] = \exp\{-bt\}s^k/(s+t)^k$; from Theorem 1, the dynamics of the total population size is governed by the recurrence equation

$$N_{t+1} = N_t \exp\{-b\}\left[\frac{s+t}{s+t+1}\right]^k \exp\left\{\frac{1}{N_t} - \frac{\gamma}{N_t^2}\right\}. \tag{7}$$

The possible dynamical behavior of the population is shown in figure 3.

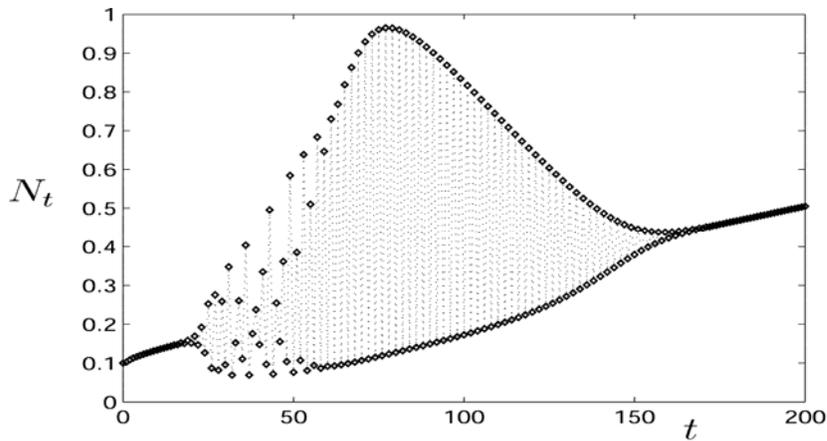

Fig. 3. The behavior of total population size $N_t$ in model (7) with $b=0$ and $\gamma = 0.046$.

Figure 3 shows that the total size of the population on its way to a stable state can experience dramatically different behavior. In particular, the trajectory may tend to a stable state, and on this way it may experience dramatically different behaviour with time from apparently chaotic oscillations to oscillatory-like changes and then to smooth changes. The "moving in the opposite direction" from the smooth changes to chaotic oscillations is also possible dependently on the particular parameter distribution and the initial values of the parameter γ and the population size.

The main peculiarity of the inhomogeneous model (7) compared to the homogeneous one, (6), is that the complex transition behavior can exist. The inhomogeneous versions of model (6) are explored in detail in [4].

**Discussion**

We have shown that inhomogeneous maps possess some essential new dynamical behaviours comparing with their homogeneous counterparts. Non-homogeneity of the population together with the natural selection lead to changing of the structure of population with time. As a result, typical trajectory of an inhomogeneous map mimics in some sense the bifurcation diagram of the corresponding homogeneous map.

It is well known that mathematical models constructed in the framework of non-linear maps can describe some surprising phenomena in behavior of biological populations (see, e.g., [5]). Most models assume that individuals within a given population are identical; equivalently, these models operate with the mean value of the reproduction rate. We have shown that modeling of inhomogeneous population dynamics on the basis of only the mean value of the reproduction rate without taking into account its distribution or at least variance is likely to be substantially incorrect. Indeed, even the dynamics of simplest inhomogeneous maps of the Malthusian type with the same initial mean value of the Malthusian parameter can be very different depending on the initial distribution of the parameter (see Theorem 2 and Example A).

The complex transition behaviors of inhomogeneous maps are the consequence of interplay of two independent factors: heterogeneity of the population and density-dependent regulatory mechanism. Let us emphasize that the evolution of the distribution of the parameter (that is, the behavior of frequencies of different type individuals) is regular and completely described by Theorem 1. In many important cases (see Theorem 2) the distribution of parameters is of the same "type" as the initial one (i.e., Gamma or Beta –distributions), but with changing in time parameters. Let us point out that all real populations are inhomogeneous.

**References.**